\documentclass[fleqn,10pt]{wlscirep}
\usepackage[utf8]{inputenc}
\usepackage[T1]{fontenc}

\title{{\small{non-peer-reviewed arXiv manuscript}}\\Long-range fiber-optic earthquake sensing by active phase noise cancellation}

\author[1]{Sebastian Noe}
\author[2]{Dominik Husmann}
\author[1]{Nils M\"{u}ller}
\author[2]{Jacques Morel}
\author[1,*]{Andreas Fichtner}
\affil[1]{Institute of Geophysics, ETH Zurich, 8092 Zurich, Switzerland}
\affil[2]{Swiss Federal Institute of Metrology, METAS, 3003 Bern-Wabern, Switzerland}

\affil[*]{andreas.fichtner@erdw.ethz.ch}

\begin{abstract}
We present a long-range fiber-optic environmental deformation sensor based on active phase noise cancellation (PNC) in metrological frequency dissemination. PNC sensing exploits recordings of a compensation frequency that is commonly discarded. Without the need for dedicated measurement devices, it operates synchronously with metrological services, suggesting that existing phase-stabilized metrological networks can be co-used effortlessly as environmental sensors. The compatibility of PNC sensing with inline amplification enables the interrogation of cables with lengths beyond 1000 km, making it a potential contributor to earthquake detection and early warning in the oceans. Using spectral-element wavefield simulations that accurately account for complex cable geometry, we compare observed and computed recordings of the compensation frequency for a magnitude 3.9 earthquake in south-eastern France and a 123 km fiber link between Bern and Basel, Switzerland. The match in both phase and amplitude indicates that PNC sensing can be used quantitatively, for example, in earthquake detection and characterization.
\end{abstract}

\begin{document}

\flushbottom
\maketitle
%
%
\thispagestyle{empty}

\section*{Introduction}

During the past decade, Distributed Acoustic Sensing (DAS) has become a mature technology that offers high spatial sampling and large frequency bandwidth from the mHz to kHz range \cite{Lindsey_2020,Paitz_2021}. It has thereby opened diverse research opportunities with immediate societal relevance, for instance, in seismic imaging and monitoring of near-surface structures and reservoirs \cite{Daley_2013,Daley_2016,Dou_2017,Martin_2017,Ajo_2019}, the detection and characterisation of volcano seismicity for potential early warning \cite{Currenti_2021,Klaasen_2021,Jousset_2022,Klaasen_2022} and studies of the structure and dynamics of glaciers and ice sheets \cite{Walter_2020,Hudson_2021,Fichtner_2022b,Fichtner_2023}. 

In sync with the popularization of DAS, novel sensing approaches have been developed to overcome two of its drawbacks: the high cost of DAS units and the maximum interrogation distance of typically several tens of kilometers. Exploiting deformation-dependent birefringence, it has been shown that optical polarization changes accumulated along transoceanic telecommunication cables record seismic ground motion \cite{Mecozzi_2021,Zhan_2021}. In an earlier study, it was demonstrated that optical phase changes in ultrastable laser signals, transmitted through metrology or telecommunication networks of hundreds to thousands of kilometers length, are sensitive to a broad range of environmental signals, including earthquakes\cite{Marra_2018}. Adopting a conceptually similar approach, a microwave frequency fiber interferometer (MFFI) has been developed at a fraction of the cost of commercial DAS units\cite{Bogris_2022}, making this technology attractive for environmental and natural hazard applications in low-income countries. A side-by-side comparison of DAS and MFFI highlighted the potential of the latter for quantitative science\cite{Bowden_2022}. While technologies based on phase transmission only provide spatially integrated, instead of distributed, deformation measurements, some level of spatial resolution can be achieved either through the use of repeaters between fiber segments\cite{Marra_2022} or a time-dependent analysis of the signals\cite{Fichtner_2022c}. Fiber-optic sensing technologies based on polarization or phase transmission greatly increase coverage, especially in the oceans, with obvious benefits for seismic imaging, as well as earthquake and tsunami early warning. However, they require dedicated measurement equipment \cite{Marra_2018,Mecozzi_2021,Bogris_2022} and possibly the interruption of the service for which a fiber is supposed to be used primarily \cite{Marra_2018}.

Here we present an alternative approach to long-range fiber-optic deformation sensing that is based on active phase noise cancellation (PNC). Commonly used to stabilize frequency dissemination in metrological fiber networks, PNC produces optical phase change measurements as a side product that is typically discarded or only monitored for system health surveillance. Through a comparison with full-waveform simulations of a regional earthquake, we demonstrate that PNC provides quantitative measurements of ground deformation without any interruption of the metrological frequency dissemination. This implies that existing metrological networks can be converted into long-range deformation sensors without additional cost and effort. 

\section*{Measurement principle and instrument response}

\begin{figure}[ht]
\centering
\includegraphics[width=\linewidth]{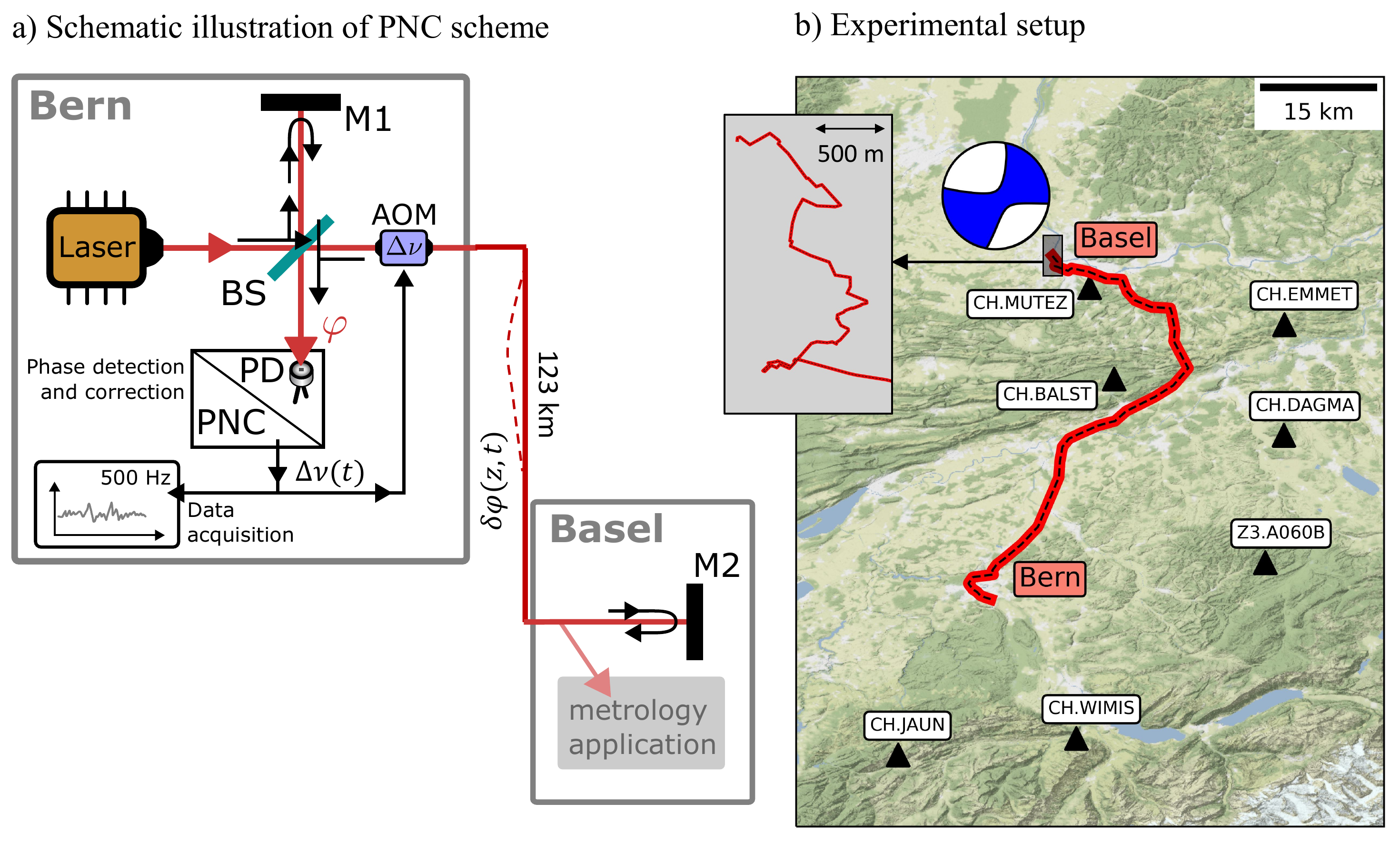}
\caption{Measurement principle and experimental setup. (a) Schematics of the interferometric phase measurement for PNC. The signal source is an ultrastable laser. A beam splitter (BS) and two mirrors (M1 and M2) form a Michelson-type interferometer that measures the optical phase noise $\varphi$ accumulated on the 123 km long interferometer arm connecting the laboratories in Bern and Basel. The optical phase is detected on a photodiode (PD) and processed to generate a correction frequency $\Delta\nu(t)$ via a phase noise cancellation setup (PNC). This frequency correction is imposed on the optical frequency using an acousto-optic modulator (AOM), thereby compensating the phase noise. We record $\Delta\nu$ with a sampling rate of 500 Hz. In Basel, part of the optical frequency is coupled out for local use in metrology applications. (b) Geometry of the fiber-optic cable connecting METAS in Bern to the University of Basel. The close-up shows the complex cable geometry within the city of Basel. The epicenter and source mechanism of the Mulhouse earthquake are marked by the beach ball. Black triangles indicate seismic stations that provided recordings for the validation of the seismic velocity model (see Fig. \ref{F:seismic}).}\label{F:setup}
\end{figure}

Phase-stabilized optical fiber networks are widely used to transmit highly stable and accurate optical frequencies from one location to another. Frequency perturbations induced by mechanical or thermal disturbances along the fiber can be compensated with the help of active PNC. Several PNC fiber networks have been developed recently, primarily with the goal to improve the performance of frequency dissemination beyond that of satellite techniques, which is a necessity for state-of-the-art atomic clock comparison \cite{Beloy_2021,Fujieda_2011,Cantin_2021,Predehl_2012,Clivati_2016,Cizek_2022,Schioppo_2022}. 
While the time-dependent frequency or phase correction applied by active PNC is \emph{per se} not a useful signal from a metrological perspective, it carries potentially valuable information on ground deformation, induced, for instance, by earthquakes. A detailed description of the PNC network utilized in this work is given in \cite{Husmann_2021}. Here we provide a condensed summary of the essentials, complemented by a schematic summary in Fig. \ref{F:setup}a.

PNC is based on a coherent optical phase measurement and feedback loop \cite{Ma_1994}, where a low-noise continuous-wave laser signal is sent from a local to a remote station through a fiber of length $L$. At the remote station, part of the signal is coupled out for local use, while the remainder is reflected back to the local station for phase detection. At some position $z$ along the fiber, an inline strain $\varepsilon(z,t)$ causes the optical phase perturbation

\begin{equation}\label{E:001}
\delta\varphi(z,t) = \frac{2\pi\nu \alpha}{c}\,\varepsilon(z,t)\,,
\end{equation}

where $\nu$ is the laser frequency and $c$ is the speed of light in the fiber \cite{Fichtner_2022c}. The coefficient $\alpha$ describes the mechanical coupling of the fiber to the solid Earth, and typically has to be inferred experimentally. It encapsulates a broad range of effects, such as the mechanical insulation of the fiber within the cable and site-effects related to unknown small-scale subsurface structure. Assuming that the maximum propagation delay $2L/c$ is small compared to the time scales of deformation, the accumulated phase change at the local station equals

\begin{equation}\label{E:002}
\varphi(t) = 2 \int\limits_{z=0}^L \delta\varphi(z,t)\, dz = \frac{4\pi\nu\alpha}{c}\,\int\limits_{z=0}^L \varepsilon(z,t)\,dz\,.
\end{equation}

The retrieved phase error is transformed into a correction frequency $\Delta\nu$, which is imposed on the optical frequency by means of an acousto-optic modulator. Provided that $\varphi(t)$ varies slowly relative to the PNC bandwidth of $c/(4L)=406$ Hz, the time derivative $\dot{\varphi}(t)$ is a good approximation of $\Delta\nu$, which constitutes our measurement quantity. Hence, in summary, we find that the instrument response between deformation $\varepsilon$ and the measured voltage $\Delta\nu$ is given by

\begin{equation}\label{E:003}
\Delta\nu(t) = \frac{4\pi\nu\alpha}{c}\,\int\limits_{z=0}^L \dot{\varepsilon}(z,t)\,dz\,.
\end{equation}

In common practice, the correction frequency $\Delta\nu$ is logged for system health monitoring or completely discarded. Here we store its time trace, thereby transforming the PNC into a deformation sensor.

\section*{Experimental setup}

We implemented the PNC system on a 123 km long fiber connecting the Swiss Federal Institute of Metrology (METAS) in Bern and the remote station at the University of Basel, as shown in Fig. \ref{F:setup}b.  This segment is part of a larger fiber network originally developed for research in precision spectroscopy \cite{Husmann_2021}, and is integrated into the telecommunication infrastructure of the Swiss national research and education network provided by SWITCH.  Integrating a phase-stabilized frequency signal into an operational fiber network requires careful consideration of the network design in order to prevent interference with co-existing spectral bands for telecommunication. A simple but expensive approach is to use dedicated dark fibers. A more cost-effective solution is provided by dark channels, where the signal is multiplexed into an unused spectral band. In contrast to other established frequency metrology networks in the C-band (1530–1565 nm wavelength), here we chose a dark channel in the L-band (1565–1625 nm wavelength) at a frequency of 190.7~THz (1572.06 nm wavelength), corresponding to ITU-T channel 7. In addition to reduced fiber lease cost, this provides a large spectral guard band separation from C-band telecommunication data traffic. 

At two points in the network, bidirectional erbium-doped fiber amplifiers are placed to compensate the optical power losses in the network. The loop bandwidth of our system, i.e., the maximum frequency at which phase noise can be compensated, is $\nu_\text{PNC}\approx 250$ Hz, slightly below the theoretical limit of $L/4c = 406$ Hz, imposed by the fiber delay. Accordingly, we measure $\Delta\nu$ with a rate of 500 samples per second, corresponding to a Nyquist frequency of 250 Hz, which is far above the frequency range of seismic events.

\section*{Observation and modelling of the 2022 M3.9 Mulhouse earthquake}

\begin{figure}[ht]
\centering
\includegraphics[width=\linewidth]{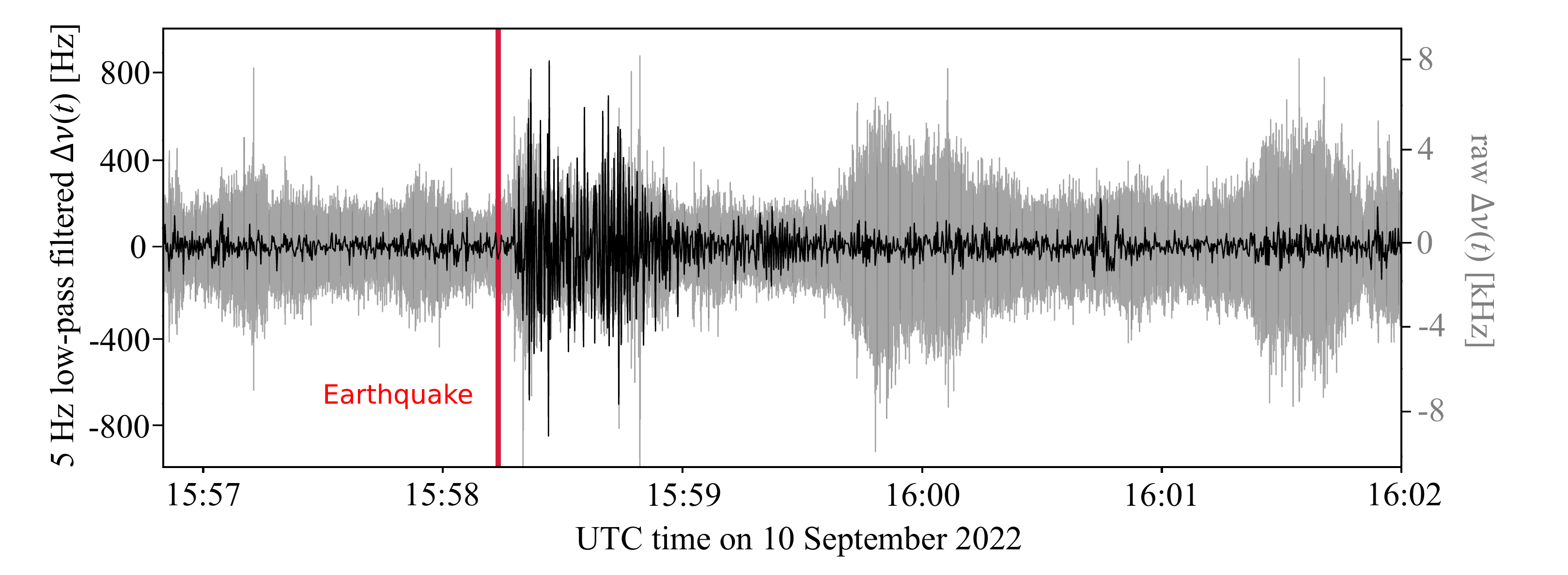}
\caption{Raw and 5 Hz low-pass filtered recordings of the PNC correction frequency $\Delta\nu(t)$. The earthquake signal dominates over the electronic and anthropogenic noise at frequencies below $\sim$5 Hz.}\label{F:recordings}
\end{figure}

On 10 September 2022, the M3.9 Mulhouse earthquake provided a unique opportunity to test the ability of the PNC system to act as a seismic sensor. Raw and low-pass filtered recordings of the correction frequency $\Delta\nu(t)$ are shown in Fig. \ref{F:recordings}. At frequencies above $\sim$30 Hz, the prevalent anthropogenic noise masks the earthquake signal. After applying a 5 Hz low-pass filter, the phase distortions caused by the earthquake is unveiled.

\begin{figure}[ht]
\centering
\includegraphics[width=\linewidth]{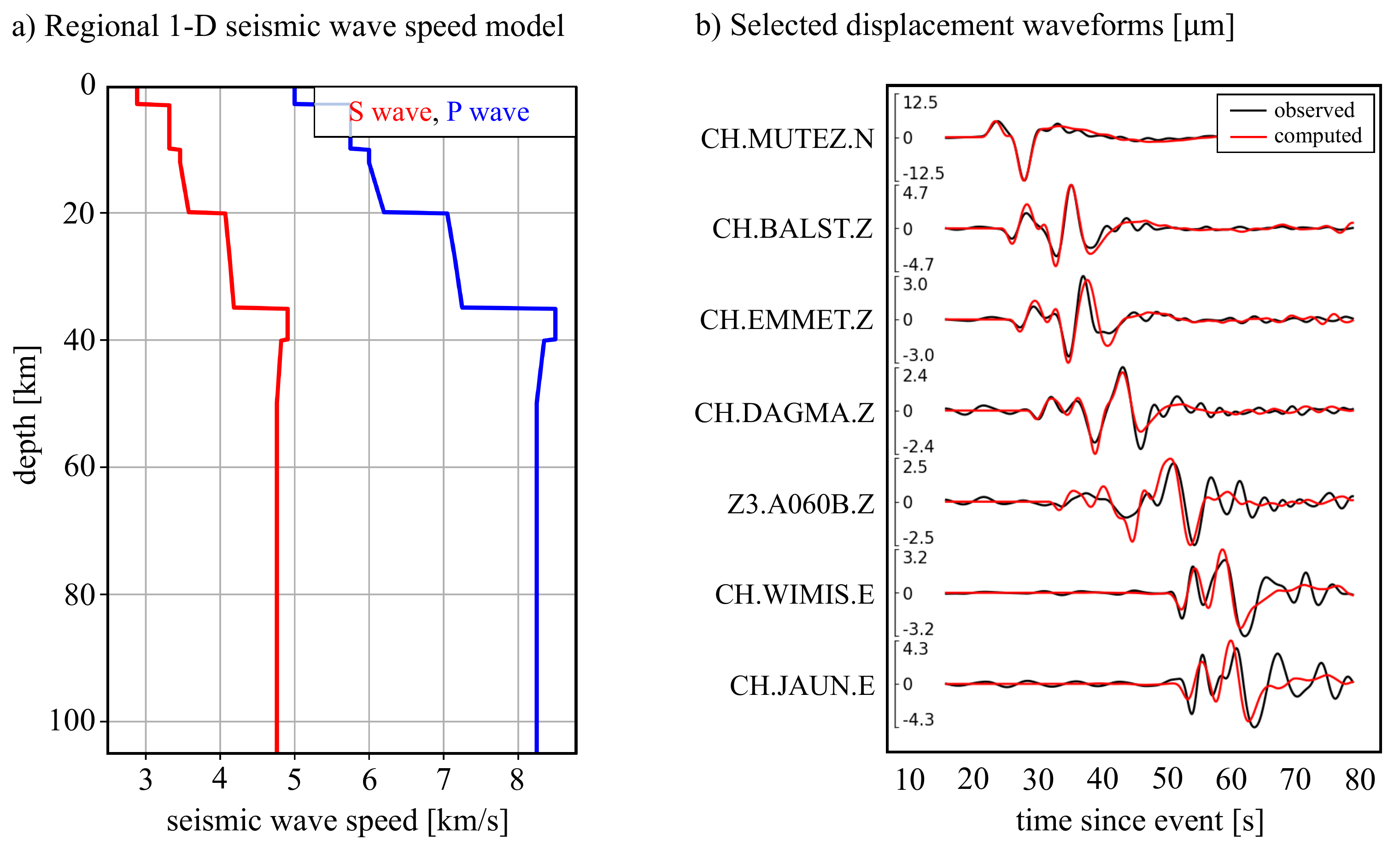}
\caption{Regional Earth model and seismogram comparison. a) Regional 1-D model of P-wave and S-wave speed, slightly modified from a seismic model of the wider Alpine region\cite{Diehl_2009} to better match observed waveforms from the Mulhouse event. b) Comparison of observed (black) and computed (red) displacement waveforms for a selection of components and stations in the Basel-Bern region.}\label{F:seismic}
\end{figure}

To assess the extent to which $\Delta\nu(t)$ may be exploited in the quantitative solution of seismological problems, we perform a comparison to simulated data. For this, we slightly modified an existing 1-D seismic velocity model of the wider Alpine region \cite{Diehl_2009}, such that it explains three-component seismometer recordings with a minimum period of 3 s in the vicinity of the cable roughly to within the noise. In this process, we did not attempt to match the fiber-optic recordings in order to avoid optimistically biased results. Fig. \ref{F:seismic}a displays the depth distributions of P- and S-wave speeds. Wavefield simulations are based on the spectral-element solver Salvus \cite{Afanasiev_2019}, which takes topographic variations into account. Earthquake source information, including location and moment tensor, are provided by GEOFON \cite{Hanka_1994}. A representative collection of waveform comparisons for different stations and components is shown in Fig. \ref{F:seismic}b. It demonstrates that the 1-D model, despite its simplicity, explains arrival times with an accuracy of $\sim\!1$ s, and amplitudes to within $\sim\!10$ \%.   

To proceed with the simulation of the correction frequency $\Delta\nu(t)$, we add the 123 km fiber-optic cable to the spectral-element mesh. Special care must be taken because transmitted phase changes strongly depend on the geometry of high-curvature segments of the cable \cite{Fichtner_2022c}. For this, we position anchor points every $\sim$50 m along rather straight cable segments, and more densely around narrow bends. In between the anchor points, the cable is represented by a cubic spline. Along the numerical cable, we output the axial strain rate $\dot{\varepsilon}(t)$ every 2 m, integrate and scale according to Eq. (\ref{E:003}), initially assuming a coupling coefficient of $\alpha=1$. 

\begin{figure}[ht]
\centering
\includegraphics[width=\linewidth]{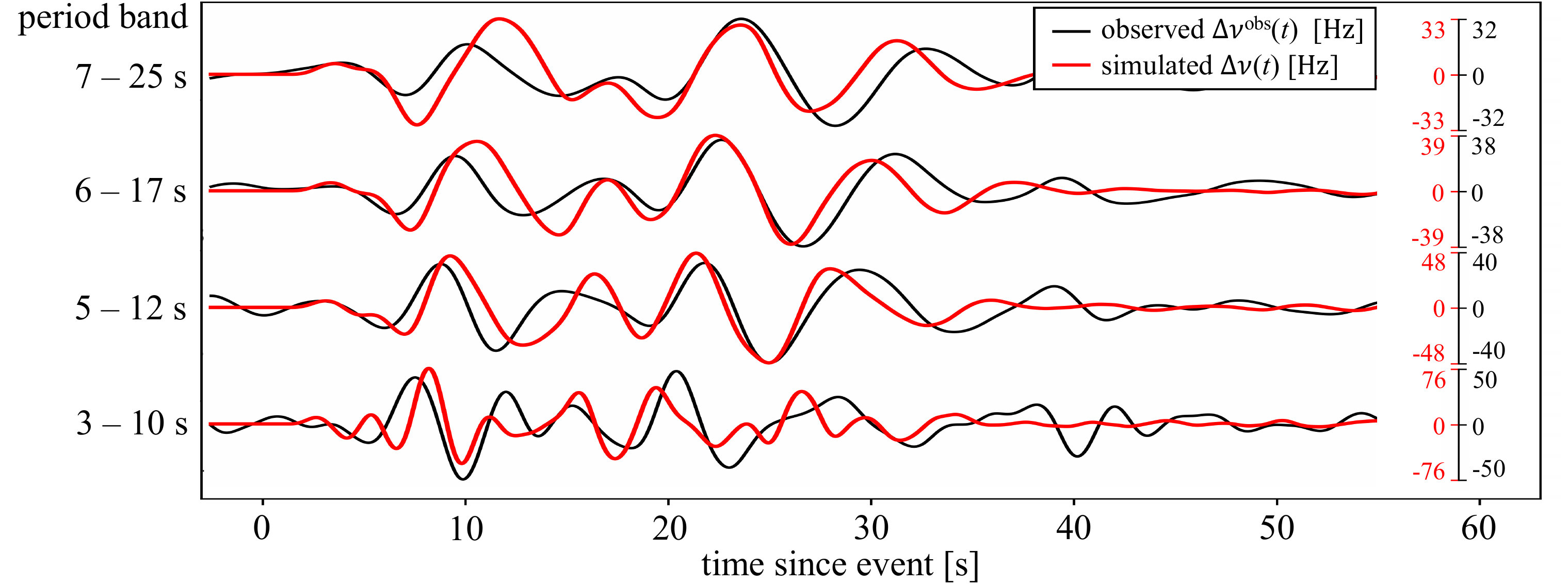}
\caption{Comparison of the observed correction frequency $\Delta\nu^\text{obs}(t)$ in black and its simulated counterpart $\Delta\nu(t)$ in red for period bands of 7 - 25 s, 6 - 17 s, 5 - 12 s and 3 - 10 s.}\label{F:comparison}
\end{figure}

A comparison of the observed frequency $\Delta\nu^\text{obs}(t)$ and its simulated counterpart $\Delta\nu(t)$ is presented in Fig. \ref{F:comparison} for different period bands, where mostly surface waves are being observed. Though waveform differences are overall larger than for the conventional displacement recordings in Fig. \ref{F:seismic}, there is a clear association between subsequent oscillation cycles, with time shifts around few seconds. Amplitude differences are dominated by the unknown $\alpha$, which Fig. \ref{F:comparison} suggests to be $\sim$0.95 in the 7 - 25 s period band, and $\sim$0.66 at lower period between 3 - 10 s. Hence, most of the earthquake-induced deformation is effectively transmitted into the fiber. Within the waveforms, time-dependent amplitude differences are in the few tens of percent range. Generally, the correction frequency time series are more complex than the displacement time series at individual seismometers because the wave interacts with the cable for a longer time; around 20-30 s. Especially the high-curvature points along the cable produce high-amplitude oscillations in the $\Delta\nu$ time series \cite{Fichtner_2022c}.

\section*{Discussion}

The major advantages of PNC-based sensing introduced here are as follows: (1) The compatibility with inline amplification permits the deployment of this scheme on fibers beyond 1000 km length. (2) Other than active PNC, already operational in many metrology networks, no dedicated measurement devices are required, meaning that there is no additional cost and effort. (3) PNC-based sensing operates without interruption of metrological services. In contrast, the metrology links used in earlier geophysical applications\cite{Marra_2018} were, to our knowledge, not stabilized during the measurement in order to directly record optical phase changes, without passing via the correction frequency. While permitting more direct phase read out, this method is not compatible with simultaneous network usage for metrological frequency dissemination. 

While previous studies \cite{Marra_2018,Marra_2022} already provided useful analyses of phase transmission signals, Fig. \ref{F:comparison} constitutes, to the best of our knowledge, the first comparison of observed and computed phase transmission waveforms. It is based on a rigorous forward modelling theory \cite{Fichtner_2022c} combined with spectral-element wavefield simulations that properly account for details of the fiber-optic cable geometry. Though slightly worse than the displacement seismogram comparison in Fig. \ref{F:seismic}b, time shifts are on the order of 1 s. This is comparable to arrival time residuals in regional seismic waveform inversion \cite{Tape_2010,Cubuk_2017}. For our experiment, the amplitude mismatch for periods above $\sim$3 s is below $\sim$66 \%. This would translate to an estimation error of the earthquake magnitude of merely $\sim$0.12. Hence, in summary, our results suggest that PNC sensing can contribute to quantitative seismological research, including the tomographic refinement of subsurface models and the characterisation of earthquakes. 

Estimated values of the coupling coefficient $\alpha$ indicate that most of the strain is actually transmitted into the fiber at periods above $\sim$3 s. Though the telecommunication cable that has not been installed for sensing applications, this result is similar to dedicated fiber-optic sensing installations where $\alpha$ can reach values close to $1$ \cite{Paitz_2021}.

This work presents a prototype application that may still be improved. Most importantly, $\alpha$ should be estimated more precisely with the help of active-source tests at various positions along the cable and with co-located seismometers. Furthermore, the actual geometry of the cable may not be known with sufficient accuracy. In fact, the geographic length of the cable, shown in Fig. \ref{F:setup}b is $\sim$8 km shorter than the cable length of 123 km that has been measured optically with high precision. Most likely, the missing $\sim$8 km have been deployed in the form of loops, which can make a significant contribution to the waveform differences in Fig. \ref{F:comparison}.

\bibliography{biblio}

\section*{Acknowledgements}

We thank the group of Prof. Stefan Willitsch and his group for continued support with the PNC installation in Basel. We acknowledge SWITCH for providing the fiber network infrastructure and its geo data, and Fabian Mauchle for technical support with the network, as well as Daniel Bowden, Sixtine Dromigny, Pascal Edme, Sara Klaasen, Patrick Paitz and Krystyna Smolinski for many fruitful discussions about this work. Funding was provided by the European Union’s Horizon 2020 research and innovation program under the Marie Sklodowska-Curie grant agreement No. 955515 (SPIN ITN), and by the Swiss National Science Foundation (SNSF) Sinergia grant CRSII5\_183579.

\section*{Author contributions statement}

D.H., S.N., A.F and J.M. conceived the experiment. D.H. conducted the experiment and collected data. A.F. and S.N. conceived the seismic data comparison, which was performed by S.N.. N.M. refined the seismic velocity model. All authors wrote and reviewed the manuscript.

\section*{Additional information}

None of the authors has any competing interests. All data analyzed during the current study are available from the corresponding author on reasonable request.

\end{document}